\documentclass[12pt]{article}
\usepackage{latexsym,epsfig,amssymb,amsmath,cite,verbatim,mcite,units}

\usepackage{amsfonts,amsthm} 
\usepackage{bm,bbm}
\usepackage{graphicx}
\usepackage{esint}
\usepackage{color}
\usepackage{braket}

\newcommand{\be}{\begin{equation}}
\newcommand{\ee}{\end{equation}}
\newcommand{\ba}{\begin{eqnarray}}
\newcommand{\ea}{\end{eqnarray}}
\def\bs{\begin{subequations}}
\def\es{\end{subequations}}
\def\a{\alpha}

\def\P{\Phi}

\def\S{S}
\def\Sb{\bar S}

\def\cD{{\cal D}}

\def\cN{{\cal N}}

\def\p{\partial}

\newcommand{\Eq}[1]{(\ref{#1})}

\def\Kahler{K\"{a}hler~}

\begin{document}

\begin{titlepage}

\begin{flushright}
\small ~~
\end{flushright}

\bigskip

\begin{center}

\vskip 2cm

{\LARGE \bf \Kahler potentials for Planck inflation} \\[6mm]

{\bf Diederik Roest, Marco Scalisi and Ivonne Zavala}\\

\vskip 25pt

{\em Centre for Theoretical Physics,\\
University of Groningen, \\
Nijenborgh 4, 9747 AG Groningen, The Netherlands\\
{\small {\tt \{d.roest, m.scalisi, e.i.zavala\}@rug.nl}}} \\

\vskip 0.8cm

\end{center}

\vskip 1cm

\begin{center} {\bf ABSTRACT}\\[3ex]

\begin{minipage}{13cm}
\small

We assess which \Kahler potentials in supergravity lead to viable single-field inflationary models that are consistent with Planck. We highlight the role of symmetries, such as shift, Heisenberg and supersymmetry, in these constructions. Also the connections to string theory are pointed out. Finally, we discuss a supergravity model for arbitrary inflationary potentials that is suitable for open string inflation and generalise it to the case of  closed string inflation. Our model includes the recently discussed supergravity reformulation of the Starobinsky model of inflation as well as an interesting alternative with comparable predictions.

\end{minipage}

\end{center}


\vfill

\end{titlepage}

\tableofcontents


\section{Introduction}

Implementing inflation into a complete high energy physics scenario remains an important challenge. Due to the high value of the potential energy of the scalar field responsible for this primordial accelerated expansion, it is natural to consider inflation in frameworks such as supergravity and string theory. Moreover, impressive progress has been made on the observational side, culminating in the recent CMB measurements by Planck \cite{Planck}. These have confirmed beyond doubt the deviation from the Harrison-Zel'dovich spectrum,
 \begin{align}
  n_s = 0.9603 \pm 0.0073 \,, \label{tilt}
 \end{align}
and have placed stronger constraints on possible non-Gaussianities and tensor modes. All observations are consistent with the `vanilla' flavour of inflation, consisting of slow-roll, single-field and small field, amidst a host of more elaborate models with clear observational signatures\footnote{See \cite{EI} for a recent comparison of the predictions of several inflationary models with Planck data.}. It therefore appears imperative to investigate possible compelling theoretical arguments for this observed simplicity.

Supersymmetry (SUSY) is another notion that seeks to unify and therefore simplify our theories of Nature. However, it is by no means apparent that supersymmetric inflationary scenarios are also necessary simple. In fact, generically the opposite appears to hold. For starters, supersymmetry naturally predicts a second quasi-light field around the Hubble scale\footnote{This is termed quasi-single field inflation in \cite{Chen}, where the inflationary consequences of an additional scalar field in the complementary series of De Sitter's unitary irreps with $0 < m^2 \leq 9/4$ were investigated.}. This is a consequence of the spontaneous breaking of supersymmetry at the inflationary Hubble scale. Any scalars other than the inflaton - of which there is always at least one, given that scalars come in pairs in SUSY models - therefore naturally acquire Hubble-scale masses. A concrete demonstration of this phenomenon can be found in \cite{Baumann}.

A second, equally generic obstacle for vanilla inflation in supergravity, is formed by the sGoldstino directions. This is a pair of scalar fields that is singled out by the spontaneous breaking of SUSY. These fields  generically correspond to unstable directions on the scalar manifold, signalling instabilities \cite{GR1, GR2, GR3}. Indeed, their inflationary consequences are dire: subject to three assumptions that we will discuss below, it can be shown that single-field, slow-roll and small field inflation cannot be realised in supergravity \cite{BRZ}. This follows from a general inequality that we will refer to as the {\it geometric bound}, as it involves the curvature on the \Kahler manifold spanned by the scalars. Note that such inflationary implications of SUSY breaking have been derived by employing the sGoldstino directions, but they hold in full generality provided the following assumptions hold. First of all, any vector multiplets play a subdominant role, that is, the effective description is F-term supergravity. Secondly, the gravitino mass has to be orders of magnitude below the inflationary scale, as also argued in \cite{Covi}. Finally, the inflaton and the sGoldstini have a non-negligible overlap: these preferred directions in field space are not (almost) orthogonal.

It is worth elaborating on the latter assumption as it has an interesting group-theoretical underpinning. As emphasised in the effective field theory approach to inflation, the inflaton can be seen as the Goldstone boson arising from the spontaneous breaking of time translation invariance: this symmetry is necessarily broken during inflation (as exemplified by \eqref{tilt}) giving rise to a Goldstone boson, which can be seen as either a scalar field or an additional helicity-0 component in the metric. This is analogous to the additional degrees of freedom of the $W^\pm$ and $Z^0$ vector bosons as arising in the Higgs mechanism. A slightly different reasoning applies to spontaneous breaking of SUSY. In this case, the Goldstone modes are a pair of spin-1/2 fermions, whose supersymmetric partners are spin-0 fields. These are referred to as Goldstini and sGoldstini fields, resp. Their emergence is completely analogous to the Higgs boson itself in the spontaneous breaking of gauge symmetry: the Higgs is the gauge partner of the aforementioned Goldstone bosons. Thus, there are interesting similarities and differences between these interpretations of the inflaton and the sGoldstini scalar fields: both arise as a consequence from the spontaneous breaking of a local symmetry (i.e.~time translational invariance and SUSY), in which they are Goldstone modes or the partners thereof.

Also from a phenomenological point of view it is interesting to focus on the third of the assumptions as a means of evading the bound of \cite{BRZ}. In fact, in the case of complete orthogonality between the inflaton and the sGoldstino, an arbitrary single-field inflationary potential can be generated in $\cN = 1$ supergravity. This model has been put forward by Kallosh, Linde and Rube (KLR) \cite{KLR} and has a very specific class of K\"{a}hler potentials $K$ and superpotentials $W$, as will be reviewed in section 2. The general structure of supergravity, therefore, does allow for the simplest case with slow-roll and small field, but only under the specific assumption of orthogonality. In any other cases, one finds that SUSY breaking poses strong constraints on the possibilities for inflation. 

As $\cN = 1$ supergravity is not a UV-complete theory of quantum gravity, it is worthwhile to investigate to what extent such a model is also applicable in string theory. In other words, is it possible to generate the required \Kahler and superpotentials in a string-theoretic configuration? We will argue that this is generically not the case. We therefore look for, and find, a relevant generalisation of the KLR model, whose form is compatible with string theory. Interestingly, this model contains the so-called Cecotti model \cite{Cecotti} that can be used to construct Starobinsky inflation in supergravity. We will moreover point out a natural and interesting generalisation of this construction, which lead to an alternative inflationary model with predictions that are comparable to Starobinsky's.

The organisation of this paper is as follows. In section 2 we review inflationary models based on a single superfield and point out the problems of these constructions. We also discuss some subtleties regarding the eta problem and the role of the inflaton, which have not been stressed in the literature. In section 3, we discuss the KLR model with two superfields as well as its (debatable) embedding in string theory. We propose a generalisation of this model in section 3.3. Finally, we end with our conclusions.

\bigskip

{\bf Note added:} upon completion of this paper we received the interesting preprint \cite{Ellis2} which has some overlap with the present discussion.


\section{sGoldstino inflation}

In this section we will discuss a number of single superfield models of inflation. In this case the inflaton necessarily coincides with the sGoldstino directions. We will consider two types of \Kahler potentials, being the shift-symmetric and logarithmic ones.

The original proposals of inflationary theories, with e.g.~hilltop inflation, were criticised for their extreme sensitivity to initial conditions. Chaotic inflation with e.g.~a quadratic potential, where inflation takes place for generic initial conditions,  was proposed to evade this problem \cite{Linde}. However, it was long unclear whether such models are strictly bottom-up and phenomenological, or whether they could also arise in top-down approaches such as supergravity and/or string theory\footnote{See e.g.~\cite{Warm,KLcaotic} for recent interesting extensions of the chaotic inflationary model.}. 

A generic problem arising in such constructions is the $\eta$-problem of supergravity: due to the specific form of the scalar potential of this class of theories, the second slow-roll parameter $\eta$ generically will receive contributions of order one \cite{Copeland}. To realise slow-roll inflation one must therefore either resort to an undesired amount of fine-tuning to cancel such terms, or eliminate these contributions altogether by means of a symmetry. The latter is termed natural inflation \cite{Freese}. It was first employed in a supergravity context to realise chaotic inflation \cite{KYY}. Instead of the canonical \Kahler potential 
 \begin{align} 
  K = \Phi \bar \Phi \,, \label{canonical-K}
 \end{align}
the authors opted for 
\begin{align}
  K = - \tfrac12 \left(\Phi - \bar \Phi\right)^2 \,. \label{shift-K}
\end{align} 
Due to the absence of the real part of the superfield $\Phi$, the \Kahler potential has a shift symmetry $\Phi \rightarrow \Phi + a$ with $a \in \mathbb{R}$.  This symmetry is the key to avoid the $\eta$-problem; relatedly, it prevents the inflaton potential from blowing up for large values of the inflaton field Re$(\Phi)$. Indeed, by recalling the expression of the scalar potential in supergravity, 
\be \label{scpot}
V= e^K \left(K^{i\bar{j}}\cD_i W\cD_{\bar{j}} \bar{W}-3|W|^2\right),
\ee
where $\cD_i W \equiv \p_iW+K_i W$, it is evident that the dangerous term $e^K$ keeps increasing exponentially in the direction Im$(\P)$ while it remains constant  in Re$(\P)$. The shift symmetry is broken only by $W$, thus generating the inflaton potential.

However, the above discussion can also be misleading as it fails to take into account the following subtlety. The careful reader may have noticed that the two quoted \Kahler potentials are in fact related by a \Kahler transformation, and hence are physically equivalent. Yet more strikingly, the \Kahler potential can even be brought to the form 
 \begin{align}
  K  = \tfrac12 \left(\Phi + \bar \Phi\right)^2 \,,
\end{align}
by means of an additional \Kahler transformation. The three forms suggest symmetry protection for either none\footnote{In fact, the canonical form \eqref{canonical-K} does not depend on the complex phase of the field $\Phi$; however, in the origin of the orthogonal, radial direction, the phase is not a physical field.}, the real or the imaginary components, respectively. How does it come about that one \Kahler potential suffers from the $\eta$-problem, while physically equivalent potentials avoid it by means of a shift symmetry, which however protects different components? The answer to this apparent conundrum is that the $\eta$-problem is not only a statement about the \Kahler potential, but also about the `naturalness' of the superpotential. For {\it generic} choices of the superpotential, one needs a shift symmetry in $K$ to keep $\eta$ small; without that shift symmetry in the \Kahler potential, one needs a carefully picked $W$ to compensate for the order-one contribution to $\eta$. These two situations can be related by a \Kahler transformations and are exactly the options alluded to  above, i.e.~fine-tuning or symmetry. Thus the form of the \Kahler potential is not the only ingredient in evading the $\eta$-problem; also the generic or fine-tuned form of the superpotential comes into play.

Following the above logic, one can try to construct an inflationary potential based on a supergravity model with a single superfield and a general superpotential. In view of the latter, we choose the shift-symmetric \Kahler potential \eqref{shift-K}. This allows for a truncation to only the imaginary part of $\Phi$ provided one takes
 \begin{align}
  W = f(\Phi) \,, \label{real-W}
 \end{align}
where the function $f$ is a real holomorphic function; in other words, when expanded in terms of its holomorphic argument, all coefficients are required to be real. The mass spectrum for this model reads
 \begin{align}
  m^2_{{\rm Re}(\Phi)} & = - 6 f'^2 - 6 f f'' + 2 f''^2 + 2 f' f''' \,, \notag \\ 
  m^2_{{\rm Im}(\Phi)} & = 4 V + 4 f^2 + 2 f'^2 -2 f f'' + 2 f''^2 -2 f' f''' \,.
 \end{align}
when evaluated at $\Phi = \bar \Phi$ and where primes denote derivatives with respect to the variables the function depends on. In this set-up, the imaginary part of the superfield $\Phi$ will generically give rise to a Hubble-scale field, which is stabilised at zero. This is exactly as expected, as the \Kahler potential contributes order one to $\eta$, and the contribution from $W$ will generically be much smaller (think e.g.~a sum of exponential - the shift symmetry of the imaginary part is mildly broken, leading to a small contribution). Hence the total $\eta$ will be order one, allowing stabilisation of this field. In contrast, the real part Re$(\P)= \phi$ will be light and play the role of the inflaton. The resulting scalar potential reads
 \begin{align}
  V = - 3 f(\phi)^2 + f'(\phi)^2 \,.
 \end{align}
This model thus allows for a truncation to a single field. However, the form of this potential is clearly not the most general due to the appearance of both the function $f$ and its derivatives. For large field inflation with a single monomial dominating the superpotential at large field values, the negative-definite contribution dominates the scalar potential. Thus, it is impossible to realise in particular chaotic inflation with a  monomial in this way.

One way of understanding the restrictions on this scalar potential, and in particular its inflationary properties, is the aforementioned geometric bound. In the case of a single superfield, the inflaton necessarily coincides with a linear combination of the sGoldstini, and hence is subject to this bound. Indeed, this set-up has been termed sGoldstino inflation, and has recently been investigated in e.g.~\cite{AlvarezGaume1, AlvarezGaume2, Ana}. The results from the latter references indicate that while large field inflation is virtually ruled out, small field inflation is only possible with severe restrictions. Two explicit examples of the latter were given, with polynomials of fourth order as superpotentials\footnote{Ref.~\cite{Ana} employed a canonical rather than shift-symmetric \Kahler potential in these examples, but this difference is immaterial as their examples involve inflection point inflation and hence almost take place at a point in moduli space. We thank Pablo Ortiz and Marieke Postma for a discussion on this point.}. We have verified that the ensuing small-field trajectories actually saturate rather than satisfy the bound of \cite{BRZ}. The resulting spectral index is $n_s = 0.92$ and hence too far on the red side of the spectrum to be compatible with Planck.

We would like to point out an additional possibility to at least ameliorate the $\eta$-problem. This arises by taking a logarithmic rather than polynomial \Kahler potential. The choice
 \begin{align}
   K = - \alpha \log \left(\Phi + \bar \Phi\right) \,,
 \end{align}
leads to a \Kahler manifold $SU(1,1) / U(1)$ whose curvature is parameterised by $\alpha$. Thus, it is well motivated from a supergravity point of view as well as from string theory, as we will discuss in the next section. In addition it eliminates the dangerous exponential terms arising from the overall \Kahler exponential in the scalar potential. Therefore, there is no longer a compelling reason to identify the imaginary part of $\Phi$, which now enjoys the shift symmetry of $K$, with the inflaton. This is good news, as it is generically {\it inconsistent} to set the real part of $\Phi$ equal to zero in order to obtain a single field model. The fact that this is consistent in the model with the shift symmetric \Kahler potential is a consequence of the square in \eqref{shift-K}. As the logarithmic \Kahler potential no longer has this feature, the only consistent truncation is to the real part. For this one needs to take the same requirement on the superpotential \eqref{real-W} being a real function of $\Phi$.

This model leads to the scalar potential:
 \begin{align}
  V = \frac{1}{\left(\Phi + \bar \Phi\right)^\alpha} \left[ (-3 + \alpha) |f|^2 + \frac{\left(\Phi + \bar \Phi\right)^2}{\alpha} |f'|^2 - \left(\Phi + \bar \Phi\right)\left( f \bar f' + \bar f f'\right)\right] \,.
 \end{align}
The choice $\alpha = 3$ is special due to the no-scale structure, where the negative definite term is exactly cancelled. With e.g.~a simple polynomial choice of the superpotential that will also arise in the next section,
 \begin{align}
  W = 3 M (\Phi -1)
 \end{align}
with $M$ a dimensionful constant, this leads to the following inflationary potential for a single scalar field $\varphi$ with canonically normalised kinetic terms:
 \begin{align}
 V = \tfrac{3}{2} M^2 e^{-\sqrt{\frac{2}{3}} \varphi}\left(-2+ e^{-\sqrt{\frac{2}{3}}\varphi}\right) \,.
 \end{align}
Therefore this potential  cannot support inflation. The only value of $\alpha$ which has a constant asympotical behaviour is $\alpha = 2$, for which the potential becomes
\begin{align}
V = - \tfrac92 M^2 e^{- \varphi} \left[2 \cosh(\varphi) + \sinh(\varphi) -3\right] \,.
\end{align}
However, in this case the negative definite contributions dominate and the scalar potential asymptotes to a negative plateau for $\varphi$ large. Again, this does not lead to a viable inflationary scenario.

\section{Orthogonal inflation}

In this section we will discuss a number of two-superfield models of inflation. A common feature of this section will be the orthogonality between the sGoldstino directions and the inflaton.

\subsection{The Kallosh-Linde-Rube model}

From the above discussion one can conclude that the choice \eqref{shift-K} suffices to evade the $\eta$-problem but not the sGoldstino bound. However, the shift-symmetric \Kahler potential is only the first ingredient that was introduced to realise chaotic inflation in supergravity. The other is a second superfield $\S$, that will be vanishing during inflation and nevertheless plays a crucial role. The pioneering model of \cite{KYY} consists of
 \begin{align}
  K = - \tfrac12 \left(\Phi - \bar \Phi \right)^2 + S \bar{S} \,, \qquad W = M S \Phi \,, \label{KYY}
 \end{align}
in terms of a real constant $M$. Inflation can be chosen to take place along $\Phi - \bar \Phi = S = 0$, while the remaining degree of freedom Re$(\P)=\phi$ has a quadratic scalar potential:
  \begin{align}
  V = M^2 \phi^2 \,.
 \end{align}
However, in this case the three truncated fields are not yet stabilised: while the imaginary part of $\Phi$ has a Hubble-scale mass in compliance with the $\eta$-problem, this is not the case for the $S$-field. To this end one can add a higher-order term to the \Kahler potential,
 \begin{align}
  K = - \tfrac12 \left(\Phi - \bar \Phi\right)^2 + S \bar{S} + \zeta  \left( S \bar{S}\right)^2 \,,
 \end{align}
which parametrises its curvature. The ensuing mass eigenvalues are
\begin{align}
 m^2_{{\rm Im}(\Phi)} = V + M^2 \,, \quad m^2_{S} = \zeta V + M^2 \,. 
\end{align}
For coefficients $\zeta $ of order one this will indeed allow both Im($\Phi)$ and $S$ to be stabilised.

At this point, it is noteworthy to focus on one general feature of this class of models. The only non-vanishing contribution to eq.~\Eq{scpot} comes from the term $\cD_S W$ that defines the sGoldstino direction in field space \cite{GR1, GR2, GR3}. This sheds light on the peculiar role of the field $S$ in the model: $S$ belongs to the sGoldstino supermultiplet and inflation happens in the {\em orthogonal} direction to the one defined by the sGoldstino along which supersymmetry is broken. The complete orthogonality of such two  directions allows to avoid both the $\eta$-problem as well as the geometric bound.

Much more recently, a new development has build on this model to generate other inflationary potentials in a similar manner, see e.g.~\cite{KL}. This has culminated in a model by Kallosh, Linde and Rube (KLR) \cite{KLR}, which consists of a prescription of how to build a class of supergravity models allowing for a completely arbitrary inflaton potential $V(\phi)$. Similar to the previous model, it consists of two complex scalar fields $\P$ and $\S$. The role of both fields will be identical to before; the real part Re$(\P) = \phi$ will be the inflaton field while Im$(\P)$ and $S$ will be essential in order to stabilise the inflationary trajectory, along which such fields will vanish. However, the \Kahler and superpotential are generalised to the following:
\be \label{KLR}
  K = K \left((\P - \bar \P)^2, \S \Sb, \S^2, \Sb^2\right), \quad W=S f(\P).
\ee
The \Kahler potential can be an arbitrary function of the arguments as indicated, and as a consequence it is separately invariant under the following transformations:
\ba
S \rightarrow -S \,, \quad \Phi\rightarrow - {\Phi}, \quad \Phi\rightarrow \Phi +a, \qquad a \in \mathbb{R} \,. \label{sym}
\ea
Similarly, $f(\P)$ is an arbitrary but real holomorphic function of the variable $\P$. 

Amongst the main novelties of such a model is that a completely general inflationary potential can be generated from a supergravity model. Moreover, given $K$ and $W$,  one does not need to perform long calculations without knowing whether the final form of the potential will be actually suitable for inflation or not. Within this model, the form of the inflaton potential will always be 
\be \label{infpot}
V(\phi)=  f(\phi)^2,
\ee
which is a completely general positive function of $\phi$. This functional freedom is guaranteed by the symmetries of the K\"ahler potential $K$ and by the linearity of $W$ in $S$. 

In the above derivation we have set the three fields that are not protected by the shift symmetry, i.e.~$S$ and $\Phi - \bar \Phi$, equal to zero. The consistency of this truncation can be seen from the full mass matrix, which gives rise to the following eigenvalues:
\begin{align}
 m^2_{{\rm Im}(\Phi)} & =  f^2 \left(1 - K_{\Phi \bar \Phi S \bar S} - \tfrac12 \partial^2_\Phi \ln(f) \right) \,, \notag \\
 m^2_{S} & =  - \left( K_{S \bar S S \bar S} \pm \left|K_{SSS \Sb} - K_{SS}\right|\right) f^2 + (\partial_\Phi f)^2 \,. 
\end{align}
Thus, for suitably chosen \Kahler manifolds with the right sectional curvature, the mass of these components is indeed Hubble-scale and hence they are stabilised at their origin.

\subsection{Embedding in string theory}

In this section we discuss to what extent the successful model of supergravity inflation with  general scalar potentials can be implemented in string theory: are there examples of \Kahler and superpotentials that follow from a string-theoretic configuration and  have the required structure\footnote{When restricting to the fields $\S$ and $\Phi$, which will generically be a subset of all fields in string-theoretic scenarios, we are assuming that this is a consistent procedure and will not address the subtleties of such truncations as pointed out in e.g.~\cite{Hardeman}.}? 

Let us start by discussing {\it open string fields} as candidates for inflation, the most famous case being D-brane inflation \cite{DbI1,DbI2,DbI3}, where the position of a D-brane in the internal compact dimensions plays the role of the inflaton. However, other open strings, such as more generic matter fields, can also be considered as inflaton candidates. Matter fields (including open string moduli) in  string theory obtain a K\"ahler potential of the form\footnote{In this subsection the fields $\Phi$ and $S$ do not necessarily denote the inflaton and the sGoldstino; instead, their role should be clear from the context.}
\ba\label{KMatter}
&& K= \alpha \Phi \bar \Phi   \qquad {\rm or } \qquad K=  \alpha \left( \Phi - \bar \Phi \right)^2  \,.  \label{KM}
\ea
Here we have assumed that any closed string moduli have been stabilised and their vevs are taken into account in the constant $\alpha$. 
Hence matter fields can satisfy all symmetry requirements \eqref{sym}. Therefore, from the point of view of the K\"ahler potential, the matter sector alone can provide sGoldstino and inflaton candidates within the KLR model. Examples of matter fields with a shift symmetry have been discussed in the context of D-brane inflation with D3/D7 in \cite{D3D7} and more recently in the context of fluxbrane inflation with D7/D7 in \cite{Hebecker}. These K\"ahler potentials can also arise for some matter fields in heterotic theory \cite{Cardoso,CWL}. Moreover, the superpotential for matter fields generically turns out to be of the form\footnote{The somewhat unconventional $i$ arises in the superpotential as a consequence of the choice of \Kahler potential \eqref{KM} depending on $\Phi - \bar \Phi$ rather than $\Phi + \bar \Phi$, as often considered in the literature.}
\be\label{WM}
W = \beta \sum_n^N \prod_{\alpha_n} i \Phi_{\alpha_n} \,,
\ee
where again, we take into account a likely dependence on any closed string moduli vevs into the constant $\beta$.
From the structure above we have the following properties:
\begin{itemize}
\item We generically expect to get the sum over several couplings for all the fields involved, including the sGoldstino, which is in contrast to the linear structure of \eqref{KLR}. 
\item On the other hand, there is a simple case which can fit completely. If one is allowed to truncate the superpotential to only a single term in the sum over $n$ in $W$ above, then it is always possible to  add a phase such that the superpotential has the form $W= S f(\Phi_i)$ with $f$ real, which fits the KLR form.
\end{itemize}
In conclusion, having matter fields alone in a configuration where some of these have a shift symmetry, it is possible, restricting to a single term in $W$, to obtain inflation from a general potential from  string theory supergravity. Extra sectors in the configuration can be added to $W$ so long a separation is possible, for example as it happened in \cite{KLOR}.

The other possibility to consider is geometric \textit{closed string moduli} in string theory. Generically these fields are the dilaton $S$, the complex structure moduli $U$ and the K\"ahler moduli $T$. These  have a well known K\"ahler potential, which takes the form
\be
 K = - \alpha \log{\left(\Phi + \bar \Phi\right)} \label{Klog}
\ee
where we denote $\Phi = \{ S, T, U \}$ and $\alpha$ here parametrises the curvature of the $SU(1,1)/$ $U(1)$ manifold. Such fields do enjoy the shift symmetry (which in this case we take in the imaginary part) of $\Phi$ but not the $\mathbb{Z}_2$ symmetry $\Phi \rightarrow - \Phi$. Therefore, while closed string moduli can potentially be identified with the inflaton sector, we cannot implement the model \eqref{KLR}. Moreover, there is no field that can be  identified with the sGoldstino direction. Turning to the superpotential:
\begin{itemize}
\item  If we consider the shift symmetry to be broken only by {\em non-perturbative} effects, the superpotential turns out to be a function of $\Phi$. This is in contrast to the requirements of the KLR model, in which $f$ would be a real function of $i \Phi$ in the conventions where the \Kahler potential has the shift symmetry in the imaginary part of $\Phi$ rather than the real part. For example, this is the case of $W \propto e^{-a \Phi}$ for $T$ in type IIB compactifications with fluxes considered widely in the literature.

\item On the other hand, if the shift symmetry is broken at tree, {\em perturbative} level, then $W$ is indeed a function of $i\Phi$, consistent with KLR. This is the case  of the tree-level superpotentials for the $S T U $-moduli generated via bulk, geometric and non-geometric fluxes.
\end{itemize}
In conclusion the closed string sector provides promising inflationary directions; however, the lack of $\mathbb{Z}_2$-symmetric \Kahler potentials prevent the implementation of the model \eqref{KLR}. 

From the discussion above, an interesting hybrid emerges naturally: the case when both matter and closed string moduli play a role. The identifications of the fields with the relevant sectors of SUSY breaking and inflation are clear: a K\"ahler modulus is identified with the inflaton sector, while a matter field is identified with the sGoldstino sector. As we have discussed, this requires a modification of the KLR model, so as to be able to relax the second condition of \eqref{sym}.
In this case generically  we expect  the K\"ahler potential to be of the form:
 \be
  K = - \alpha  \log \left( \Phi + \bar \Phi - S \bar S\right) \,, \label{log-K}
 \ee
where $\Phi$ is identified with a closed string modulus, for example the K\"ahler modulus, and $S$ is identified with some matter field, for example a brane position. The curvature parameter $\alpha$ can take the values $1,2,3$. Further, the superpotential for $\Phi$ has to come from non-perturbative contributions, since this is how geometric moduli and matter fields couple. We will discuss the details of such a model with arbitrary superpotentials and a \Kahler potential of the form \eqref{log-K} in the next subsection.


\subsection{General string-inspired inflationary potentials}

We now turn to our model which does not require the $\mathbb{Z}_2$ symmetry for the field $\Phi$. We do, however, retain the shift symmetry for $\Phi$, as well as the $\mathbb{Z}_2$ symmetry for the field $S$. Moreover, the superpotential is identical to the previous model. Therefore, we have the following specifications:
 \begin{align}
   K = K \left( \Phi + \bar \Phi, S \bar S, S^2, \bar S^2\right), \quad W = S f(\Phi) \,,
\end{align}
where $f$ is still a real holomorphic function. This model allows for a consistent truncation to the inflationary trajectory at Im$(\Phi) = S = 0$.

Note that an important difference with respect to the KLR model is that the shift symmetry no longer coincides with the inflaton direction. In our case, the inflaton is still identified with the real part of $\Phi$ and hence does appear in the \Kahler potential. This is necessary to allow for a consistent truncation (we comment on the orthogonal identification in the discussion), but does reintroduce the $\eta$-problem. Thus, we will have to carefully choose $K$ and $f$ in order to end up with a light inflaton field. However, we will see that a logarithmic \Kahler potential at least ameliorates the $\eta$-problem, in the sense of eliminating exponential in favour of polynomial contributions to the scalar potential.

In the truncation to a single real scalar field $\Phi = \phi$, the scalar potential reads
 \begin{align}
  V = e^{K} K^{S \bar S} f(\phi)^2 \,, \label{general-V}
 \end{align}
evaluated at the inflationary trajectory Im$(\Phi) = S = 0$. Due to the functional form of $V$, this model also allows for a general inflationary potential for the single field $\phi$ that we are retaining. The consistency of this truncation is guaranteed by construction: the field equations of the three truncated fields are satisfied at the inflationary trajectory. However, this does not imply that this truncation is also stable. For this, one needs to consider the mass spectrum of such fields. In order for effective single-field behaviour, these will have to be super-Hubble. We will later check, in explicit examples, to what extent this condition can be met.

An important class of the models considered here depends only on the specific combination 
 \begin{align}
  X = \Phi + \bar \Phi - S \bar S \,.
 \end{align}
In this case the shift symmetry is enhanced to the three-dimensional Heisenberg group, which acts in the following way
 \begin{align}
  \Phi \rightarrow \Phi + i a + \bar b S + \tfrac12 |b|^2 \,, \quad S \rightarrow S + b  \,, \quad a \in \mathbb{R}\,, b \in \mathbb{C} \,. \label{Heisenberg}
 \end{align}
Although generally applicable, we will concentrate on \Kahler potentials with the following dependence on the Heisenberg invariant:
 \begin{align}
 K = - \alpha \log (X) \,. \label{Heisenberg-K}
 \end{align}
This type of potential is very natural in both string theory and supergravity. The corresponding \Kahler manifold is $SU(2,1) / U(2)$. 

With this choice, the scalar potential becomes:
\ba\label{VHeis}
V&=& \frac{|S|^2}{\a}\left(  X^{1-\a}|S|^2 + X^{2-\a} \right)\left| f' + \frac{ \a f }{X}\right|^2  + \frac{X^{1-\a}}{\a}|f|^2\left(1+ \frac{\a |S|^2}{X}\right)^2 \nonumber \\
&&  \hskip-0.2cm 
+\frac{X^{1-\a}}{\a}|S|^2\left(1+\frac{\a |S|^2}{X}\right)\left[  \bar f \left(f' + \frac{\a f}{X} \right) +   f \left(\bar f' + \frac{ \a \bar f }{X}\right)  \right] - 3\frac{|S|^2|f|^2}{X^\a}\,.
\ea
Thus at $S=0$ the potential is simply
\be
V= \frac{X^{1-\alpha}|f|^2}{\a}.
\ee
In terms of real and imaginary parts for $\Phi$ and the field $S$, the masses of the fields read:
\begin{align}
m_{{\rm Re}(\Phi)}^2 & = \frac{2}{\a} \left[X^{1-\a}\left(\a -2+\frac{1}{\a}\right) f^2 +  X^{2-\a}\left(\frac{3}{\a} -2\right)f f'  + \frac{X^{3-\a}}{2\a} (f f'' + f'^2)\right], \notag \\
m_{{\rm Im}(\Phi)}^2 & = \frac{2}{\a} \left[X^{1-\a}\left(1-\frac{1}{\a}\right) f^2 -  \frac{X^{2-\a}}{\a} f f'  + \frac{X^{3-\a}}{2\a} (f'^2-f f'')\right], \notag \\
m_S^2 & = \frac{X^{1-\a}}{\a} \left(\a -2-\frac{1}{\a}\right) f^2 + \frac{X^{2-\a}}{\a} \left(\frac{2}{\a}-2\right) f'f + \frac{X^{3-\a}}{\a^2} f'^2. \label{spectrum}
\end{align}
Clearly, for a successful single-field inflationary model, the first of these has to be light whereas the latter three degrees of freedom need to be stabilised, either around or above the Hubble scale.

An example of such a model was recently discussed in connection with the Starobinsky model of inflation\footnote{
A similar set-up with an identical \Kahler potential \eqref{Starobinsky} was also recently used to embed the Starobinsky model in supergravity \cite{Ellis}. However, in that case the inflaton was identified with one of the directions of $S$, while the $\Phi$ field was argued to be stabilised by other means. Hence, the relation to the present model is unclear at present.} \cite{SC, Staro1,Staro2}. Specifically, it employs the Cecotti model \cite{Cecotti} with the following potentials:
 \begin{align}
  K = - 3 \log \left( \Phi + \bar \Phi - S \bar S\right) \,, \quad
  W = 3 M S (\Phi -1) \,. \label{Starobinsky}
 \end{align}
In terms of a canonically normalised scalar field $\varphi$, this yields the scalar potential
 \begin{align}
  V = \tfrac34 M^2 \left( 1 -  e^{ - \sqrt{2} \varphi / \sqrt{3}} \right)^2 \,.
 \end{align}
Inflation takes place at large $\varphi$. In this limit the three masses become
 \begin{align}
  m^2 = \{ 0, 4 H^2, -2 H^2 \} \,,
 \end{align}
where $H^2 = V/3$. This has been demonstrated to be equivalent to Starobinsky's $R + R^2$ model of inflation \cite{Starobinsky}. The predictions of this model are in excellent agreement with the Planck data. The relations of this model to superconformal supergravity have been discussed in \cite{SC}. In this reference it was also been pointed out that the $S$ field is not stable with this \Kahler choice;  to this end one could add a stabilising term $ \beta (S \bar S)^2 / (\Phi + \bar\Phi)$ to the argument of the logarithm leading to
 \begin{align}
  m_S^2 = (-2 + 4 \beta)H^2\,,
 \end{align}
which is finite along the whole inflationary trajectory and positive for an appropriate choice of $\beta$. Instead, we find that the imaginary part of $\Phi$ is stable with the present \Kahler potential and hence poses no problems for inflation.

A generalisation of the Starobinsky model arises when one allows for an arbitrary curvature of the \Kahler manifold, i.e.~including the parameter $\alpha$. For simplicity we will keep the same superpotential. Following the same line of reasoning, one ends up with a scalar potential for a canonically normalised scalar field $\varphi$ that reads
 \begin{align}
   V = \frac{2^{1 - \alpha} (3M)^2}{\alpha} \Big[ e^{  (3 - \a) \varphi / \sqrt{2 \a}} - e^{(1 - \a ) \varphi / \sqrt{2 \a}} \Big]^2 \,.
 \end{align}
For generic values of $\alpha$ this will lead to an exponential potential for large $|\varphi|$. There are only two exceptions to this behaviour: the first is for $\alpha = 3$ discussed above, while the second is for $\a = 1$. Interestingly, this value is also consistent with string theory and leads to inflation for large and negative $\varphi$. The scalar potential becomes
 \begin{align}
  V = 9 M^2 \left( e^{\sqrt{2} \varphi} - 1\right)^2 \,,
 \end{align}
and thus can be obtained from the Starobinsky potential by a sign flip and stretching in the $\varphi$ direction. Nevertheless, in this case one needs to stabilise even along the $Im\Phi$ direction as the three masses \eqref{spectrum} asymptote to
 \begin{align}
  m^2 = \{ 0,  0, -6 H^2 \} \,. 
 \end{align}
Having a viable single-field scenario translates into adding a term $-\gamma S\bar S (\Phi-\bar\Phi)^2 / (\Phi + \bar\Phi)^2$, together with the same term stabilising $S$ in the case $\a=3$, to the argument of the logarithm. With these choices, the mass spectrum turns to be finite along the inflaton direction and takes the following values:
\begin{align}
  m^2 = \{ 0,  12\gamma H^2 , (-6+12\beta) H^2 \} \,. 
 \end{align}
Interestingly, a value of $\beta>1/2$ leads to positive mass of the field $S$, independently of the parameter $\a$. Moreover, this model leads to the following spectral index and tensor-to-scalar ratio for different numbers of e-foldings:
 \begin{align}
  & N = 50: \qquad n_s = 0.961 \,, \quad r = 0.0015 \,, \notag \\
  & N = 60: \qquad n_s = 0.967 \,, \quad r = 0.0011 \,.
 \end{align}
Comparable to Starobinsky's, these are also comfortably consistent with the Planck results.

\section{Discussion}\label{disc}

In this paper we have assessed a number of functional forms of the \Kahler potentials that allow for truncation to a single scalar field that is identified as the inflaton. In order to circumvent the $\eta$-problem, either a shift symmetric or a logarithmic \Kahler potential was employed. Moreover, in order to circumvent the geometric bound, a second superfield $S$ was introduced. The superpotential was taken to be linear in this field, leading to the identification of $S$ as the sGoldstino field. With a shift-symmetric \Kahler potential one can obtain an arbitrary single-field scalar potential \cite{KLR}. However, string-theoretic considerations have led us to generalise this model. In particular, we have argued that a logarithmic \Kahler potential that includes the inflaton field is equally suited to address these issues. It also leads to a general single-field inflationary potential \eqref{general-V} and allows  for stabilisation of the other degrees of freedom.

It is striking that the considerations regarding the Heisenberg invariant \Kahler potential \eqref{Heisenberg-K}  lead to two unique choices $\alpha = 3$ and $1$ that give rise to inflation. Both of these are compatible with string-theoretic arguments as well as Planck values for the spectral index and tensor-to-scalar ratio.  The former of these was recently pointed out to require stabilising terms just for the sGoldstino field while we find that the latter needs an additional term in the \Kahler in order to give a super-Hubble mass also to the imaginary part of $\Phi$. It would be interesting to investigate the connection of this choice to the purely geometric Starobinsky model $R + R^2$ and the superconformal approach. Finally, a further generalisation would involve different superpotentials to that of \eqref{Starobinsky}.

A number of interesting connections are worth pointing out. First of all, the symmetry of the Heisenberg invariant \Kahler potential has also been employed to solve the $\eta$-problem in \cite{Antusch1, Antusch2}. However, that scenario differs in an important way from the present: in that case, the field $S$ is identified as the inflaton, whereas a third superfield is added to play to the role of the sGoldstino. The inflationary predictions of that set-up are thus unrelated to ours. Secondly, one can envisage a curvaton-like scenario \cite{curvaton} with the present model, where e.g. the field $S$ would be light and generate the isocurvature fluctuations. Such a scenario with the sGoldstino field as curvaton has been explored in \cite{supercurvaton}. In the present model, this would require e.g.~a higher-order term $(S \bar S)^2$ in the argument of the logarithmic \Kahler potential with exactly the right coefficient to yield a light sGoldstino.

Finally, we would like to return to the issue of what happens to the \Kahler potential $K=K(X)$ if we identify the inflaton, in the more obvious fashion, with the imaginary part of $\Phi$.  When we set the derivative of the scalar potential with respect to Re$(\Phi)$ equal to zero, we obtain
 \be
  2  \left( \frac{K''}{K'} -K'\right) = \frac{1}{|f|^2}\left(|f^2|_{,\Phi} + |f^2|_{,\bar \Phi}  \right)\,,
 \ee
where primes denote derivatives with respect to $X$. If we now consider $f$ to be a real function of $i\Phi$, then the right-hand side of this equation vanishes at Re$(\Phi)=0$. Solving for the left-hand side then gives rise to a specific form of K\"ahler potential: the most general solution is
 \be\label{HeisK1}
  K(X) = - \log{\left( X + a\right)} + b\,,
 \ee
where $a$ and $b$ are two integration constants. Thus, the form of the K\"ahler metric is fixed to be a $SU(2,1) / U(2)$ coset manifold with a specific curvature, corresponding to $\alpha =1$. 

However, in order to truncate consistently the orthogonal direction to the inflaton, Re$(\Phi)$, one has to ensure that its equation of motion is satisfied.  It can be checked that this receives contributions from the Christoffel symbols, which do not allow us to decouple the field. However, the extra factors obtained are proportional to the slow roll parameters and, therefore, the inflationary trajectory occurs approximately along the imaginary part of $\Phi$. Similarly, if one would calculate the mass matrix for the field $\Phi$ along the putative inflationary trajectory Re$(\Phi) = S =0$, one finds that this matrix is not diagonal; in other words, the real and imaginary components are not mass eigenstates. Again, the extra factors are proportional to the slow roll parameter $\epsilon$  and, therefore, they are approximate eigenvectors during inflation. We thus find an interesting situation with pros and cons. The advantage of this truncation is that the inflaton field is naturally light, as it is protected by the shift symmetry as well as the logarithmic dependence. On the other hand, the proposed inflationary trajectory along Im$(\Phi)$ is only an approximate solution, with slow-roll suppressed deviations in the direction Re$(\Phi)$. It would be interesting to investigate whether this approximate truncation contains viable inflationary scenarios.


\section*{Acknowledgments}

{We acknowledge very stimulating discussions with Mar Bastero-Gil, Andrea Borghese, Thorsten Battefeld, Renata Kallosh, Andrei Linde, Pablo Ortiz and Marieke Postma on various aspects discussed in this paper.}

\providecommand{\href}[2]{#2}\begingroup\raggedright\endgroup

\end{document}